# 1. Introduction

The drift-tearing mode [1] involves the combined effects of magnetic reconnection and electron density and temperature gradients, has an intrinsic frequency of oscillation related to these gradients, and is characterized by both a slower growth rate and significantly different eigenfunctions when compared to the purely resistive tearing mode [2,3]. The mode, that is driven primarily by the longitudinal current density gradient, is strongly connected with the transport of electron thermal energy, and can lead to the formation of relatively large magnetic islands. The relevant excitation threshold is the same as that of the resistive tearing mode, as long as the relevant electron equation of state is adiabatic. However, the mode was found to be practically impossible to excite in collisionless regimes [4] as a result of the combined effects of electron Landau damping and temperature gradient [4] or of the equivalent effects of parallel longitudinal electron conductivity and temperature gradient [5] in weakly collisional regimes. On the other hand, as experiments with lower degrees of collisionality have been undertaken [6] modes of this kind, that produce magnetic reconnection, have been observed to persist together with the formation of the resulting macroscopic magnetic islands.

In order to resolve this paradox we consider [7] that "mesoscopic reconnecting" modes develop from a background of "micro-reconnecting" modes with short scale distances ($\lesssim c/\omega_{pe}$) that generate a series of strings of small magnetic islands and are driven by the electron temperature gradient. The envisioned effects of the latter modes are to produce a significant increase of the ratio of the transverse thermal conductivity to the longitudinal conductivity. This is shown to restore the excitation of mesoscopic mode driven mostly by the current density gradient and involving the combined effects of finite resistivity, electron thermal conductivities, and temperature gradients.

The present paper is organized as follows: In Section 2 a description is given of the micro-reconnecting modes that are driven by the electron temperature gradient and produce a sequence of small magnetic islands. Their excitation is shown to lead to an increase of the ratio of the transverse to the longitudinal electron thermal conductivity. In Section 3 the key role that the electron thermal energy balance equation has on the characteristics of the weakly collisional mesoscopic mode is demonstrated and the intermediate asymptotic region, out of the three regions that the mode involves, is identified. In Section 4, the combination of the longitudinal electron momentum conservation equation and of the quasi-neutrality condition leads to the identification of the innermost region where the singularity characteristic of the present problem is removed. This singularity concerns the function describing the difference between the relative electron temperature fluctuation and the relative plasma displacement. In Section 5 the equation characterizing the innermost region, where the effect of the magnetic diffusion coefficient is important, is given and the limits of its validity are discussed. In Section 6 the solution of the innermost region equation is discussed and integrals of it that enter the expression for the mode growth rate are estimated. In Section 7 the two characteristic components of the growth rate are derived from the matching condition of the asymptotic solution for the perturbed radial magnetic field over the three regions. The main component of the growth rate is associated with the current density gradient and the other with the temperature gradient. In Section 8 comments concerning the relevance of the mesoscopic mode to the formation of the experimentally observed magnetic islands are given.

## 2. Micro-reconnecting Modes and Their Role

In view of extending our results to more complex configurations we refer to a plane geometry where the magnetic field around a surface $x = x_0$ is represented as

$$\mathbf{B} \simeq B_z(x)\mathbf{e}_z + (x-x_0)B'_y \mathbf{e}_y. \tag{1}$$

The mesoscopic mode that we consider is envisioned to develop from a background of excited microscopic modes that produce magnetic islands on smaller scales and are localized over distances not exceeding $c/\omega_{pe}$ around successive surfaces $x = x_j$ contained in a finite interval centered on $x = x_0$.

The micro-reconnecting modes are collisionless and driven by the electron temperature gradient. The relevant perturbed longitudinal electric field $\hat{E}_\parallel$ has both $\nabla_\parallel \hat{\Phi}$ and $(1/c)\partial \hat{A}_\parallel/\partial t$ as significant components while $\hat{\mathbf{E}}_\perp \simeq -\nabla_\perp \hat{\Phi}$. In particular we take $\hat{A}_\parallel = \tilde{A}_\parallel(\Delta x_j)\exp(-i\omega t + ik_y y)$ where $\Delta x_j \equiv x - x_j$ and $\hat{A}_\parallel = \tilde{A}_\parallel(\Delta x_j)$ is an even function of $\Delta x_j$. The frozen-in condition is broken by the effects of finite electron inertia. Only $T_{e\parallel}$ is involved in these modes as we consider $k_\perp^2 \rho_e^2 \ll k_\perp^2 d_e^2$ where $k_\perp^2 = k_y^2 - \partial^2/\partial x^2$. Moreover $\omega^2 \gtrsim k_\parallel^2 \mathrm{v}_{the}^2$ for $\mathrm{v}_{the}^2 = 2T_e/m_e$ and $\omega^2 < k_\perp^2 \mathrm{v}_{thi}^2$ where $\mathrm{v}_{thi}^2 = 2T_i/m_i$ and $k_\perp^2 \rho_i^2 \gg 1$. Here $\rho_e$ and $\rho_i$ indicate the electron and the ion gyroradius, respectively. We note that $\omega = \omega_R + i\omega_I$ and that the mode frequency depends strongly on the values of $\eta_e = (d\ln T_{e\parallel}/dx)/(d\ln n/dx)$ in the sense that $\omega_R/\omega_I$ is finite for $\eta_e \gg 1$ and can vanish if $\eta_e > 1$ but not too large. The equation for $\tilde{A}_\parallel$ that we consider, in the fluid limit where $kd_e > 1$, $\omega < \omega_{*Te}$, and $\omega^2 > k_\parallel^2 \mathrm{v}_{the}^2$, is based on the collisionless longitudinal electron momentum conservation equation

$$m_e \frac{\partial}{\partial t} \hat{u}_{e\parallel} = -\frac{\mathbf{B}}{B} \cdot \nabla \hat{T}_e - \frac{\hat{B}_x}{B} \frac{dT_e}{dx} - e\hat{E}_\parallel \qquad (2)$$

where we consider $\eta_e \gg 1$. We combine Eq. (2) with the electron thermal energy balance equation $\partial \hat{T}_e/\partial t + \hat{v}_{Ex} dT_e/dx = 0$, the electron mass conservation, and the relevant Maxwell's equations to obtain

$$\left[\omega^3 + \omega_{*Te} k_y^2 c_{se}^2 \frac{(\Delta x_j)^2}{L_s^2}\right]\left(d_e^2 \frac{d^2}{dx^2} - k_y^2 d_e^2\right)\tilde{A}_\parallel(x) + \omega^2 \omega_{*Te} \tilde{A}_\parallel(x) \simeq 0, \qquad (3)$$

as $\hat{n}_i \simeq \hat{n}_e \simeq -ne\hat{\Phi}/T_i \simeq (k_\parallel \hat{u}_{e\parallel}/\omega)n$. Here $\hat{u}_{e\parallel} = -\hat{J}_\parallel/(ne)$ is the longitudinal electron flow velocity, $\hat{J}_\parallel = -c/(4\pi)\left[\partial^2/\partial x^2 - k^2\right]\hat{A}_\parallel$, $k_\parallel \simeq k_y B_y'(\Delta x_j)/B$ where $B_y'/B \equiv 1/L_s$ defines the "shearing distance", $\hat{V}_{Ex} = \hat{E}_y c/B$, $\omega_{*Te} \equiv -k_y c/(eB) dT_e/dx$ is the frequency representing the mode driving factor and $c_{se} \equiv (T_i/m_e)^{1/2}$ is the "electron sound" velocity. The boundary conditions are $\tilde{A}_\parallel = 1$ and $d\tilde{A}_\parallel/dx = 0$ for $x = 0$ and $\tilde{A}_\parallel \to 0$ for $x^2 \to \infty$. Therefore we can derive the following quadratic form, from the Fourier transform of Eq. (3),

$$\bar{\omega}^3 \left\langle |\tilde{Y}_k|^2 \right\rangle = -\left\langle \left|\frac{d}{d\bar{k}}\tilde{Y}_k\right|^2 \right\rangle + \bar{\omega}^2 \left\langle \frac{|\tilde{Y}_k|^2}{\bar{k}_0^2 + \bar{k}^2} \right\rangle \qquad (4)$$

where $\bar{\omega} \equiv \bar{k}_0^{-1/2} C^{-1/4} (L_s/c_{se})\omega$, $\bar{k}_0 \equiv (k_y d_e)^{2/3} C^{1/6}$, $C \equiv 2T_i r_{Te}^2/(\beta_e T_e L_s^2)$, $\beta_e \equiv 8\pi n T_e/B^2$, $r_{Te} \equiv |T_e/(d\ln T_e/dx)|$, $\tilde{Y}_k \equiv \tilde{A}_k (\bar{k}^2 + \bar{k}_0^2)$, $\tilde{A}_k$ is the Fourier transform of $\tilde{A}_\parallel(x)$ and $\langle \rangle = \int d\bar{k}$. This shows that there is a marginally stable mode for $\bar{k}_0 < \bar{k}_{0crit}$ and an unstable mode for $\bar{k}_0 > \bar{k}_{0crit}$. The relevant phase velocity is in the direction of the electron diamagnetic velocity.

The string of magnetic islands produced by this mode around the surface $x = x_j$ is represented by the magnetic surface function

$$\psi_j \propto (x - x_j)^2 + \delta_{Is}^2 \frac{\tilde{A}_\parallel(x - x_j)}{\tilde{A}_\parallel(0)} \cos(\omega_R t - k_y y) = c_j \qquad (5)$$

with $c_j = \text{const.}$, $\delta_{Is} \equiv (\tilde{A}_\parallel(0)/B_y')^{1/2} \exp(\omega_I t/2)$ and $\tilde{A}_\parallel(0) = \tilde{B}_{x0}/k_y$. Clearly the linear theory is valid for $\delta_{Is} \sim |\tilde{B}_{x0}/\bar{B}_y|^{1/2} (d_e r_{Te})^{1/2} < d_e$ where $\bar{B}_y \sim |B_y'| r_{Te}$. In particular, we envision that a sequence of these strings of islands is produced over a macroscopic scale distance represented by $|B_y'/B_y''|$. These strings of islands can be a hindrance for the propagation of temperature along the field and can modify, considerably, mode-particle resonances that are relevant to the excitation of the drift-tearing mode in the collisionless regime [4].

The effective transverse thermal diffusion coefficient that we associate with the micro-reconnecting modes, on the basis of the relevant quasi-linear theory [8], can be expected to be the order of $D_\perp^e \sim \alpha_d (d_e/r_{Te}) cT_e/(eB)$ where $\alpha_d$ is a numerical coefficient, $d_e = c/\omega_{pe}$, and $1/r_{Te} \equiv -d\ln T_e/dx$ and we consider the ratio $\Delta_{th} \equiv D_\perp^e/D_\parallel^e$ to be increased further relative to the classical value by the expected reduction of the longitudinal thermal diffusivity $D_\parallel^e$ resulting from the excitation of the same modes.

We observe that the well known electrostatic modes, so called ETG modes, involving smaller scale lengths than the micro-reconnecting modes can also be excited. Mode packets (quasi-modes) can be constructed [9] by a proper superposition of elementary electrostatic modes that are sharply localized in the $x$-direction like those found for ITG modes [9]. The quasi-modes can cover a macroscopic interval in the $x$-direction and can maintain temperature profiles that are localized along the magnetic field. In particular, the expression for the electron temperature is given by

$$\hat{\tilde{T}}_e \simeq \tilde{T}_{e0} W(x) \exp\left[-i\omega_R t + ik_y\left(y - x\frac{z}{L_s}\right) + \frac{i}{2}\frac{z^2 k_y^2 \sigma_I}{L_s^2 |\sigma|^2}\right] \times \exp\left[-\frac{1}{2}\frac{z^2 k_y^2}{L_s^2 |\sigma|^2}\sigma_R + \omega_I t\right] \quad (6)$$

where $W(x)$ is a "weight" function localized over a macroscopic distance, elementary modes represented by $\tilde{T}(x) = \tilde{T}_{e0}\exp(-\sigma(\Delta x_j)^2/2)$ are considered, and $\sigma = \sigma_R + i\sigma_I$. Thus, we se that $\hat{\tilde{T}}_e$ is localized along the magnetic field over the scale distance $L_z \sim L_s |\sigma|/(k_y\sqrt{\sigma_R})$ where $|\sigma|k_y^2 \sim 1$. This can be seen as a further contribution toward having a state of reduced effective longitudinal thermal conductivity.

A more complete mode dispersion equation that does not have the limitations of the fluid limit in which Eq. (2) is derived and include mode particle resonances, has been derived by the so called drift approximation. In this case, the quadratic form replacing the one that can be obtained from Eq. (3) without taking its Fourier transform is

$$d_e^2 \left\langle \left|\frac{d\tilde{A}_\parallel}{dx}\right|^2 + k^2 |\tilde{A}_\parallel|^2 \right\rangle \simeq \left\langle |\tilde{A}_\parallel|^2 \frac{\omega^2 \omega_{*Te} \Im_0^0}{\omega^3 + k_\parallel^2 c_{se}^2 \omega_{*Te} \Im_0^0}\right\rangle. \quad (7)$$

Here $\langle \; \rangle \equiv \int dx$,

$$\Im_0^0 \equiv \frac{1}{n}\int_{-\infty}^{+\infty} dv_\| F_{Me}(v_\|^2)\left(\frac{m_e v_\|^2}{T_e}\right)\frac{\mathcal{L}(v_\|^2)\bar{\bar{\omega}}^2 - \bar{\bar{\omega}}^3}{\bar{\bar{\omega}}^2 - k_\|^2 v_\|^2/\omega_{*Te}^2}, \tag{8}$$

$F_{Me} = \left[1/\left(\sqrt{\pi}v_{the}\right)\right]n\exp\left(-v_\|^2/v_{the}^2\right)$, $\bar{\bar{\omega}} \equiv \omega/\omega_{*Te}$ and $\mathcal{L}(v_\|^2) = v_\|^2/v_{the}^2 - 1/2$. It is clear that the fluid limit corresponds to $\Im_0^0 \simeq 1$.

Recently, a parallel, independent analysis of "long wavelength ($k\rho_e < 0.1$) electron temperature gradient driven modes" [10] has been announced but not presented. Therefore, a specific comparison of this analysis with the conclusions given in this section cannot be made.

## 3. Mesoscopic Mode and Relevant Electron Thermal Energy Balance Equation

The perturbed magnetic field, for the mesoscopic mode, is represented by $\hat{\mathbf{B}} = \tilde{\mathbf{B}}(x)\exp\left(-i\omega t + ik_y y\right)$. The theory of this mode involves the analysis of three asymptotic regions of which the outer (macroscopic) region involves scale distances of the order of the radius $a$ of the plasma column and in particular $1/k_y \sim a$. In this region the "hyperconductivity" condition $\hat{\mathbf{E}} + \hat{\mathbf{v}}\times\mathbf{B}/c = 0$ is held to be valid and the perturbed magnetic field is described by the quasi-neutrality condition $\nabla\cdot\mathbf{J})_{tot} \simeq \nabla_\| J_\|)_{tot} = 0$, as the effects of finite ion inertia and Larmor radius can be neglected. The relevant equation

$$\tilde{B}_x \frac{d}{dx}J_\| + (\mathbf{k}\cdot\mathbf{B})\left(\frac{d^2 \tilde{B}_x}{dx^2} - k^2 \tilde{B}_x\right) \simeq 0 \tag{9}$$

when solved over the plasma radius, leads to find a solution for $\tilde{B}_x(x)$ that is continuous at $x = 0$ but has a singular curvature $\left(d^2\tilde{B}_x/dx^2\right)$. Thus we argue that if $\delta \ll a$ represents the width of the transition region over which the singularity is removed

$$\tilde{B}_x \simeq \tilde{B}_{x0}\left[1 + \varepsilon_\delta \varphi(x/\delta)\right] \tag{10}$$

where $\varepsilon_\delta \equiv \delta/a$ and $\varphi$ is a finite function of $x/\delta$. Moreover, if we define the displacement $\hat{\boldsymbol{\xi}}$ as $\hat{\mathbf{V}} = -i\omega\hat{\boldsymbol{\xi}}$ the hyperconductivity condition gives $\hat{\xi}_x = \hat{B}_x/(i\mathbf{k}\cdot\mathbf{B})$ that is also singular at $x = 0$.

In the transition region, the relevant longitudinal momentum conservation equation is

$$0 \simeq -v_{ei}^\| nm_e\hat{u}_{e\|} - en\hat{E}_\| - \mathbf{B}\cdot\left(\nabla p_e + \alpha_T n\nabla T_e\right)/\mathrm{B} - \hat{\mathbf{B}}\cdot\left(\nabla\hat{p}_e + \alpha_T n\nabla\hat{T}_e\right)/\mathrm{B}, \tag{11}$$

where $\alpha_T$ is the thermal force coefficient and the other terms have standard definitions. The adopted electron thermal energy balance equation is

$$\frac{3}{2}n\left(\frac{\partial \hat{T}_e}{\partial t}+\hat{V}_{Ex}\frac{dT_e}{dx}\right)+nT_e\nabla_\parallel \hat{u}_{e\parallel} \simeq -\nabla\cdot\left(\frac{\mathbf{B}}{\mathrm{B}}\hat{q}_{e\parallel}+\frac{\hat{\mathbf{B}}}{\mathrm{B}}q_{e\parallel}\right)-\nabla_\perp\cdot\hat{\mathbf{q}}_{e\perp}, \quad (12)$$

where $\hat{V}_{Ex} \equiv c\hat{E}_y/B \simeq -ick_y\hat{\Phi}/B$ and we define $\hat{\tilde{\xi}}_{Ex} \equiv i\hat{V}_{Ex}/\omega$. Moreover,

$$\nabla_\perp\cdot\hat{\mathbf{q}}_{e\perp} \simeq -D_\perp^e\left(\partial^2\hat{T}_e/\partial x^2\right)3n/2,$$

$$-\nabla\cdot\left(\hat{\mathbf{B}}q_{e\parallel}/B+\mathbf{B}\hat{q}_{e\parallel}/B\right)\simeq ik_\parallel\left\{-D_\parallel^e\left[ik_\parallel\hat{T}_e+\left(\hat{B}_x/B\right)(dT_e)/dx\right]\right\}$$

and $k_\parallel \simeq k_y B_y'(x-x_0)/B \equiv k_y(x-x_0)/L_s$.

We note that the drift-tearing mode was found originally [1] in the (adiabatic) limit where the terms in the electron thermal energy balance due to the thermal conductivities are not important, while for the mesoscopic mode that we consider these terms are prevalent. Moreover, we observe that it is not contradictory to envision weakly collisional mesoscopic modes excited from a background of collisionless microscopic modes as the relevant frequencies and growth rates have very different values. In fact, the typical frequency of both modes is closely related to $\omega_*^T \equiv k_y cT_e/(eBr_{Te})$. Therefore we require that the mean free path $\lambda_{ei} = v_{the}/\nu_{ei}$ be in the following interval

$$\frac{1}{\beta_e^{1/2}} < \frac{\lambda_{ei}}{2r_{Te}} < \frac{a}{\rho_e}\frac{|\omega_*^T|}{\gamma} \quad (13)$$

that corresponds to a realistic range of plasma parameters. We note also that while in the case of microscopic modes $\gamma \equiv \mathrm{Im}\,\omega \gtrsim \mathrm{Re}\,\omega$, in the case of the mesoscopic modes $\gamma \ll |\mathrm{Re}\,\omega| \sim \omega_*^T$.

We observe that Eq. (12) reduces to

$$\left(\tilde{T}_e+\tilde{\xi}_{Ex}T_e'\right)\left(1+ik_\parallel^2\frac{D_\parallel^e}{\omega}\right)+k_\parallel^2\frac{D_\parallel^e}{\omega}\frac{\Delta\tilde{B}_x}{\mathbf{k}\cdot\mathbf{B}}T_e'-i\frac{D_\perp^e}{\omega}\frac{d^2\tilde{T}_e}{dx^2} \simeq \frac{2}{3}T_e\frac{k_\parallel\hat{u}_{e\parallel}}{\omega}, \quad (14)$$

where $T_e' \equiv dT_e/dx$ and

$$\Delta\tilde{B}_x = \tilde{B}_x - i\left(\mathbf{k}\cdot\mathbf{B}\right)\tilde{\xi}_{Ex}. \quad (15)$$

Thus in the outer region, where $\left|\Delta\tilde{B}_x\right| \ll \left|\tilde{B}_x\right|$ and the contributions to Eq. (14) associated with $D_\perp^e$ and $\tilde{u}_{e\parallel}$ are not significant,

$$\tilde{T}_e \simeq -\tilde{\xi}_{Ex} T'_e \tag{16}$$

Then we introduce the following variables

$$\tilde{L} \equiv \frac{\tilde{T}_e}{T'_e} \quad \text{and} \quad \tilde{\mathcal{L}} \equiv \tilde{\xi}_{Ex} + \tilde{L} \tag{17}$$

and rewrite Eq. (14), for the transition region, as

$$\frac{D^e_\perp}{D^e_\parallel} \frac{d^2}{dx^2} \tilde{L} - k^2_\parallel \tilde{L} \simeq -ik_\parallel \left( \frac{\tilde{B}_{x0}}{B} - \frac{2}{3} \frac{T_e}{T'_e} \frac{\tilde{u}_{e\parallel}}{D^e_\parallel} \right) - i \frac{\omega}{D^e_\parallel} \tilde{\mathcal{L}}, \tag{18}$$

for $\tilde{B}_x \simeq \tilde{B}_{x0}$. Thus we identify the intermediate region by the scale distance

$$\delta_I \equiv \Delta^{1/4}_{th} \left( \frac{L_s}{k} \right)^{1/2}, \tag{19}$$

where $\Delta_{th} \equiv D^e_\perp / D^e_\parallel$, and rewrite Eq. (18), considering $\left| \tilde{B}_{xo} \right| > \left| B \tilde{u}_{e\parallel} (T_e/T'_e) / D^e_\parallel \right|$,

$$\frac{d^2 \bar{T}}{d\bar{x}^2} - \bar{x}^2 \bar{T} = \bar{x} - \left( \frac{i\omega L^2_s}{D^e_\parallel} \right) \bar{\mathcal{L}}, \tag{20}$$

where $\bar{\mathcal{L}} \equiv (k\delta_I) i\tilde{\mathcal{L}} / (\tilde{B}_{x0}/B'_y)$, $\bar{x} \equiv x/\delta_I$ and $\bar{T} \equiv (k\delta_I) \tilde{L} / (\tilde{B}_{x0}/B'_y)$. If we define $\bar{Y} \equiv -i\tilde{\xi}_{Ex} (k\delta_I) B'_y / \tilde{B}_{x0}$ we have

$$\bar{\mathcal{L}} = \bar{T} - \bar{Y} \quad \text{and} \quad \Delta \tilde{B}_x \simeq \tilde{B}_{x0} \left[ 1 - \bar{x} \bar{Y} \right]. \tag{21}$$

As we shall show, $\left| \bar{\mathcal{L}} \right| \ll \left| \bar{T} \right|$, $\bar{T} \simeq \bar{Y}$ in the $\delta_I$-region and we consider $\left| \omega L^2_s \bar{\mathcal{L}} / D^e_\parallel \right| \ll 1$. In this case the equation for the temperature fluctuation $\tilde{T}_e$ decouples from the other equations. In particular $\bar{Y}$ is given by

$$\frac{d^2 \bar{Y}}{d\bar{x}^2} - \bar{x}^2 \bar{Y} \simeq -\bar{x} \tag{22}$$

that yields $\bar{Y} \simeq 1/\bar{x}$ for $\bar{x}^2 \gg 1$ as required by the asymptotic connection with the solution in the outer region. We note that an approximate solution of Eq. (14-III) which reproduces the asymptotic limits of the actual solution for $\bar{x}^2 \ll 1$ and $\bar{x}^2 \gg 1$ is $\bar{Y} \simeq \bar{\alpha}\bar{x}(1+\sigma_0 \bar{x}^4)/(1+\beta_0 \bar{x}^2 + \bar{\alpha}\sigma_0 \bar{x}^6)$ where $\sigma_0 \simeq 0.1$, $\bar{\alpha} \simeq 0.6$, $\beta_0 \simeq 0.4$.

## 4. Electron and Total Momentum Conservation Equations

After taking the $x$-component of $\partial \hat{\mathbf{B}}/\partial t = -c\nabla \times \hat{\mathbf{E}}$ where $\hat{\mathbf{E}} = \hat{\mathbf{E}}_\perp + \hat{E}_\parallel \mathbf{B}/B$, $\hat{\mathbf{E}}_\perp = -\hat{\mathbf{V}}_E \times \mathbf{B}/c$ and $\hat{E}_\parallel$ is given by Eq. (11), we arrive at the following equation

$$(\delta\omega)\Delta\tilde{B}_x \simeq i(\mathbf{k}\cdot\mathbf{B})\omega_*^{TT}\tilde{\mathcal{L}} + \left(iD_m - \frac{k_\parallel^2 v_{se}^2}{\omega_{*\parallel}^T}\frac{c^2}{\omega_{pe}^2}\right)\frac{d^2\tilde{B}_x}{dx^2} \qquad (23)$$

where $\tilde{\mathcal{L}}$ was defined in Section 3, $D_m \equiv v_{ei}^\parallel c^2/\omega_{pe}^2$, $v_{se}^2 \equiv T_e/m_e$, $\omega_*^{TT} \equiv \omega_{*T}(1+\alpha_T)$, $\omega_{*\parallel}^T \equiv \omega_{*e} + \omega_*^{TT}$, $\delta\omega = \omega - \omega_{*\parallel}^T$, $\omega_{*T} = -k_c(c/eB)dT_e/dx$, and we consider $|\delta\omega| \ll |\omega_{*\parallel}^T|$.

In order to complete the set of equations to be solved that include Eqs. (20) and (23) in particular, we consider the transition region that can be derived from the quasi neutrality condition $\widehat{\nabla_\perp \cdot \mathbf{J}_\perp} + \widehat{\nabla_\parallel J_\parallel} \simeq 0$ using the guiding center description. Therefore

$$-\widehat{\nabla\cdot\mathbf{J}_\perp} \simeq \widehat{\nabla_\parallel J_\parallel} \simeq \frac{\hat{B}_x}{B}\frac{d}{dx}J_\parallel + ik_\parallel \hat{J}_\parallel \simeq \frac{c}{4\pi}\left[\frac{\hat{B}_x}{B}\left(\frac{d^2}{dx^2}B_y\right) + \frac{k_\parallel}{k_y}\frac{\partial^2 \hat{B}}{\partial \hat{x}^2}\right]. \qquad (24)$$

We note that in the transition region the parity of $\tilde{B}_x$ and $d^2\tilde{B}_x/dx^2$ that is of interest in order to arrive at the expression of the growth rate is even in $x$. For this reason and in order to simplify the analysis, which would require [3,11] a separate expansion of $\tilde{B}_x$ in $\varepsilon\ln\varepsilon$, where $\varepsilon$ is the small parameter, typical of the problem under consideration, we choose to consider the case where the term involving $dJ_\parallel/dx$ can be neglected around $x=0$.

Moreover

$$\widehat{\nabla\cdot\mathbf{J}_\perp} \simeq \left.\frac{\partial}{\partial x}\hat{J}_x\right)_i \simeq ne\frac{\partial}{\partial x}\left(\hat{v}_p + \hat{v}_{FLR}\right)_{ix}. \qquad (25)$$

where $\hat{v}_p$ and $\hat{v}_{FLR}$ are the polarization and the finite ion Larmor radius drifts, respectively. In particular

$$\left(\hat{v}_p + \hat{v}_{FLR}\right)_{ix} \simeq \frac{c}{B}(-i\omega + i\omega_{di})\frac{1}{\Omega_{ci}}\left(-\frac{\partial}{\partial x}\hat{\Phi}\right) \qquad (26)$$

where $\omega_{di} = k_y \hat{v}_{di}$ and $\hat{v}_{di} = c/(eBn)dp_i/dx$ is the ion diamagnetic velocity. Then Eq. (24) reduces to

$$-i\frac{d^2\tilde{B}_x}{dx^2} \simeq \frac{4\pi n m_i}{(\mathbf{k}\cdot\mathbf{B})}\bar{\bar{\omega}}_i^2 \frac{d^2\tilde{\xi}_{Ex}}{dx^2} \ , \tag{27}$$

where $\bar{\bar{\omega}}_i^2 \equiv \omega_{*\parallel}^T(\omega_{*\parallel}^T - \omega_{di})$. In fact this can be derived also from the total momentum conservation equation [1]. Therefore Eq. (23) becomes

$$(\delta\omega)\Delta\tilde{B}_x \simeq i(k_\parallel B)\omega_*^{TT}\tilde{\tilde{\mathcal{L}}} + iD_m \frac{d^2\tilde{B}_x}{dx^2} \ , \tag{28}$$

where

$$\tilde{\tilde{\mathcal{L}}} \equiv \tilde{\mathcal{L}} - \rho_s^2 \frac{d^2\tilde{\xi}_{Ex}}{dx^2} \frac{\omega_{*\parallel}^T - \omega_{di}}{\omega_*^{TT}} \ , \tag{29}$$

and $\rho_s^2 \equiv (T_e/m_i)/\Omega_{ci}^2$. Consequently,

$$\tilde{\xi}_{Ex} = -\tilde{L} + \tilde{\tilde{\mathcal{L}}} + \rho_s^2 \frac{d^2\tilde{\xi}_{Ex}}{dx^2}\frac{\omega_{*\parallel}^T - \omega_{di}}{\omega_*^{TT}} \tag{30}$$

and considering Eq. (28)

$$\tilde{\tilde{\mathcal{L}}} \sim (\delta\omega/\omega_*^{TT})\tilde{\xi}_{Ex} \ . \tag{31}$$

Thus $\tilde{L} \simeq -\tilde{\xi}_{Ex}$ in the $\delta_I$ - region as we require that $\Delta_{th}^{1/2} > \rho_s^2 k/L_s$. If, as we shall show, $\left|(D_m/\tilde{B}_{x0})d^2\tilde{B}_x/dx^2\right| < |\delta\omega|$ in this region, we have

$$\tilde{\tilde{\mathcal{L}}} \simeq -i\frac{\delta\omega}{\omega_*^{TT}}\left(\frac{\tilde{B}_{xo}}{k_y x/L_s} - \tilde{L}\right) . \tag{32}$$

Therefore $\tilde{\tilde{\mathcal{L}}}$ tends to become singular for $\bar{x} \to 0$ and the consideration of an innermost region where this singularity is removed is necessary.

## 5. Innermost Asymptotic Region Acknowledgments

We observe that Eq. (27) can be rewritten as

$$-i\frac{d^2\tilde{B}_x}{dx^2} \simeq B\frac{\bar{\bar{\omega}}_i^2}{v_A^2}\frac{1}{k_\parallel}\frac{d^2}{dx^2}\left[\tilde{\tilde{\mathcal{L}}} - \left(1 + \alpha_0 \rho_s^2 \frac{d^2}{dx^2}\right)\tilde{L}\right] \tag{33}$$

where $\alpha_0 \equiv (\omega_{*\parallel}^T - \omega_{di})/\omega_*^{TT}$. Therefore,

$$-i\frac{d^2\tilde{B}_x}{dx^2} \simeq B\frac{\bar{\bar{\omega}}_i^2}{v_A^2} \frac{1}{k_\parallel} \left\{ \frac{d^2\tilde{\mathcal{L}}}{dx^2} - \frac{1}{\Delta_{th}}\left(k_\parallel^2 \tilde{L} - ik_\parallel \frac{\tilde{B}_{x0}}{B}\right) - \frac{\alpha_0}{\Delta_{th}}\rho_s^2 \frac{d^2}{dx^2}\left(k_\parallel^2 \tilde{L} - ik_\parallel \frac{\tilde{B}_{x0}}{B}\right) \right\}$$

and

$$-i\frac{d^2\tilde{B}_x}{dx^2} \simeq i\frac{\bar{\bar{\omega}}_i^2}{\Delta_{th} v_A^2}\left(\tilde{B}_{x0} + ik_\parallel B\tilde{L}\right) + \frac{B\bar{\bar{\omega}}_i^2}{v_A^2}\frac{1}{k_\parallel}\frac{d^2}{dx^2}\left\{\tilde{L} - \frac{\alpha_0}{\Delta_{th}}\rho_s^2 k_\parallel^2 \tilde{L}\right\}. \quad (34)$$

Then Eq. (28) becomes

$$-i\left(\delta\omega + i\left|\omega_{*\parallel}^T\right|\frac{D_\parallel^e D_m}{D_\perp^e D_A}\right)\Delta\tilde{B}_x \simeq -\left(k_\parallel B\right)\omega_*^{TT}\tilde{\bar{\mathcal{L}}} + i\left|\omega_{*\parallel}^T\right|\frac{D_m}{D_A}\frac{B}{k_\parallel}\frac{d^2}{dx^2}\left(\tilde{\bar{\mathcal{L}}} - \frac{\alpha_0}{\Delta_{th}}\rho_s^2 k_\parallel^2 \tilde{L}\right) \quad (35)$$

and this can be rewritten as

$$-i\left(\delta\omega + i\left|\omega_{*\parallel}^T\right|\varepsilon_I\right)\frac{\Delta\tilde{B}_x}{\tilde{B}_{x0}} \simeq -i\bar{x}\bar{\bar{\mathcal{L}}}\omega_*^{TT} + \left|\omega_{*\parallel}^T\right|\varepsilon_I \frac{1}{\bar{x}}\frac{d^2}{d\bar{x}^2}\left[\bar{\bar{\mathcal{L}}} - \frac{\alpha_0}{\Delta_{th}}\left(\frac{k}{L_s}\delta_I\rho_s\right)^2 \bar{x}^2\bar{T}\right]. \quad (36)$$

Here $\bar{\bar{\mathcal{L}}} \equiv \bar{\mathcal{L}} + \alpha_0\left(\rho_s^2/\delta_I^2\right)d^2\bar{Y}/d\bar{x}^2$, $D_A \equiv v_A^2/\left|\omega_{*\parallel}^T - \omega_{di}\right|$, $\varepsilon_I \equiv \varepsilon_*/\Delta_{th}$ and we consider

$$\varepsilon_* \equiv \frac{D_m}{D_A} \ll \frac{D_\perp^e}{D_\parallel^e} \equiv \Delta_{th} . \quad (37)$$

Thus we see, from Eq. (36), that an innermost layer whose width is of the order of

$$\delta_c \equiv \varepsilon_m^{1/4}/k = \varepsilon_*^{1/4}\left(\frac{L_s}{k}\right)^{1/2} < \delta_I \quad (38)$$

is to be considered. This is the layer in which the effects of finite resistivity becomes important, and we note that the adopted two-fluid description requires that $\delta_c > \rho_i$, $\rho_i$ being the ion gyro radius. Clearly, the small parameter $\varepsilon_m$ is the most appropriate to characterize the innermost region and its definition can be rewritten as

$$\varepsilon_m = \frac{D_m k^2 \left|\omega_{*\|}^T - \omega_{di}\right|}{\omega_A^2} \sim \frac{\nu_{ei}}{\Omega_{ce}} \frac{k\rho_s^2}{L_p} (kL_s)^2 \, , \tag{39}$$

where, $\omega_A^2 = v_A^2/L_s^2$, $\Omega_{ce} = eB/(m_e c)$, $\rho_s^2 = T_e/(m_i \Omega_{ci}^2)$ and $1/r_p \equiv -d\ln p_e/dx$. Then we require $\nu_{ei}/\Omega_{ce} > (\rho_i^2/L_s^2)(kr_p)T_i/T_e$ that gives a realistic condition on the range of temperatures allowed by the present theory. We observe that the condition $\nu_{ei} > |\omega|$ corresponds to $\nu_{ei}/\Omega_{ce} > k\rho_e^2/r_p$.

## 6. Solution for the Innermost Region

In the innermost region

$$\tilde{\tilde{\mathcal{L}}} \sim \frac{\delta\omega}{\omega_*^{TT}(k\delta_c)} \frac{\tilde{B}_{x0}}{B_y'} \tag{40}$$

and referring to the r.h.s. of Eq. (13) we have, in the same region,

$$k_\| \frac{\tilde{B}_{x0}}{B} \sim k\frac{\delta_c}{L_s} \frac{\tilde{B}_{x0}}{B} \quad \text{and} \quad \frac{\omega_{**}^T}{D_\|^e} \tilde{\mathcal{L}} \sim \frac{\delta\omega}{D_\|^e} \frac{L_s}{\delta_c k} \frac{\tilde{B}_{x0}}{B} \, . \quad \text{Therefore the last two terms}$$

compare as

$$\varepsilon_*^{1/2} \text{ compares to } \frac{\delta\omega L_s}{D_\|^e k} \tag{41}$$

As we shall verify *a posteriori* $\delta\omega L_s/kD_\|^e$ is in fact smaller than $\varepsilon_*^{1/2}$. Therefore $\bar{T}$ can be evaluated independently of $\bar{\mathcal{L}}$ from Eq. (20) and Eq. (36) reduces to

$$\frac{\delta\omega}{\omega_\|^T} + i\varepsilon_I \simeq \left(\bar{x}\alpha_*^T + i\varepsilon_I \frac{1}{\bar{x}}\frac{d^2}{d\bar{x}^2}\right)\bar{\bar{\mathcal{L}}} \, , \tag{42}$$

where $\alpha_*^T \equiv \omega_*^{TT}/\left|\omega_{*\|}^T\right|$, in the innermost region, for $k^2\rho_i^2 \, \delta_c^\rho/(L_s^2 \delta_I) < 1$. If we define

$$\bar{\mathcal{L}}_0 \equiv \frac{\left|\omega_{*\|}^T\right|}{\delta\omega + i\left|\omega_{*\|}^T\right|\varepsilon_I} \bar{\bar{\mathcal{L}}} \, , \tag{43}$$

Eq. (42) becomes

$$1 + \alpha_*^T \bar{x}\bar{\mathcal{L}}_0 + i\frac{\varepsilon_I}{\bar{x}}\frac{d^2\bar{\mathcal{L}}_0}{d\bar{x}^2} = 0 \, . \tag{44}$$

Furthermore, it is convenient to introduce the finite variables $\bar{x}_c \equiv x/\delta_c$ and $\bar{\mathcal{L}}_c \equiv \delta_c \left( \bar{\mathcal{L}}_0 / \delta_I \right)$ as Eq. (44) reduces to

$$1 + \alpha_*^T \bar{x}_c \bar{\mathcal{L}}_c + \frac{i}{\bar{x}_c} \frac{d^2 \bar{\mathcal{L}}_c}{d\bar{x}_c^2} = 0 . \tag{45}$$

It is easy to see that for $\bar{x}_c^2 \gg 1$, $\bar{\mathcal{L}}_c \simeq -1/\left( \alpha_*^T \bar{x}_c \right) + i 2/\left( \alpha_*^{T2} \bar{x}_c^5 \right)$ indicating that $\operatorname{Im} \bar{\mathcal{L}}_c \equiv \bar{\mathcal{L}}_{cI}$ is a more localized function of $\bar{x}_c$ than $\operatorname{Re} \bar{\mathcal{L}}_c \equiv \bar{\mathcal{L}}_{cR}$. More generally Eq. (45) can be separated into

$$1 + \alpha_*^T \bar{x}_c \bar{\mathcal{L}}_{cR} + \frac{1}{\alpha_*^T} \frac{1}{\bar{x}_c} \frac{d^2}{d\bar{x}_c^2} \left( \frac{1}{\bar{x}_c^2} \frac{d^2 \bar{\mathcal{L}}_{cR}}{d\bar{x}_c^2} \right) , \tag{46}$$

showing that the sign of $\alpha_*^T \bar{\mathcal{L}}_{cR}$ is independent of $\operatorname{sgn} \alpha_*^T$ while

$$\bar{\mathcal{L}}_{cI} = -\frac{1}{\alpha_*^T \bar{x}_c^2} \frac{d^2}{d\bar{x}^2} \bar{\mathcal{L}}_{cR} \tag{47}$$

indicates that the sign $\bar{\mathcal{L}}_{cI}$ is independent of that of $\alpha_*^T$. In particular, in the next section we shall make use of the fact that the integral

$$\int d\bar{x}_c \left( \bar{x}_c \bar{\mathcal{L}}_{cI} \right) = -\int d\bar{x}_c \left[ \left( \frac{d}{d\bar{x}_c} \bar{\mathcal{L}}_{cR} \right)^2 + \left( \frac{d}{d\bar{x}_c} \bar{\mathcal{L}}_{cI} \right)^2 \right] \tag{48}$$

is negative, and that, since

$$\int d\bar{x}_c \left[ \bar{x}_c^2 \left( 1 + \alpha_*^T \bar{x}_c \bar{\mathcal{L}}_{cR} \right) \right] = 0 , \tag{49}$$

the integral

$$\int d\bar{x}_c \left[ 1 + \alpha_*^T \bar{x}_c \bar{\mathcal{L}}_{cR} \right]$$

is positive.

### 7. Asymptotic Matching and Growth Rates

Matching the solution of the outer region to that of the inner regions corresponds to taking

$$\Delta' \equiv \frac{1}{\tilde{B}_{x0}} \left\{ \left. \frac{d\tilde{B}_x}{dx} \right|_{x \to 0+} - \left. \frac{d\tilde{B}_x}{dx} \right|_{x \to 0-} \right\} = \frac{1}{\tilde{B}_{x0}} \int_{\delta_i} dx \frac{d^2 \tilde{B}_x}{dx^2} \tag{50}$$

where $\Delta'$ is evaluated from the solution in the outer region and the integral in Eq. (50) is taken over the $\delta_I$-region. Thus integrating Eq. (33) over the same region, we have

$$-i\delta\omega \int_{-\infty}^{+\infty} d\overline{x}\left(\Delta\tilde{B}_x\right) \simeq \omega_*^{TT} \int_{-\infty}^{+\infty} d\overline{x}\tilde{\tilde{\mathcal{L}}}\left(k_\parallel B\right) + \left(\frac{D_m}{\delta_I}\Delta'\right)\tilde{B}_{x0}. \tag{51}$$

This indicates clearly that the growth rate has two components. One depending on $\Delta'$ being $\neq 0$ and positive, as will be shown, and the other depending on $\omega_{**}^T \neq 0$. If we use the finite variable $\overline{\mathcal{L}}_0$ defined by Eq. (43), Eq. (52) can be rewritten as

$$(-i\delta\omega)\int_{-\infty}^{+\infty} d\overline{x}\left\{1+\overline{x}\overline{Y}+\alpha_*^T\overline{\mathcal{L}}_0\overline{x}\right\}+\frac{\varepsilon_*}{\Delta_{th}}\omega_*^{TT}\int_{-\infty}^{+\infty} d\overline{x}\left(\overline{\mathcal{L}}_0\overline{x}\right) \simeq D_m\frac{\Delta'}{\delta_I}. \tag{52}$$

Then we make use of Eq. (44) and obtain

$$(-i\delta\omega)\int_{-\infty}^{+\infty} d\overline{x}_c\left\{1+\overline{x}\left(\alpha_*^T\overline{\mathcal{L}}_0-\overline{Y}\right)\right\} \simeq $$

$$D_m\frac{\Delta'k}{\varepsilon_m^{1/4}} + \left|\omega_{*\parallel}^T\right|\left(\frac{\varepsilon_*}{\Delta_{th}}\right)^{3/4}\int_{-\infty}^{+\infty}\left(1-\overline{x}\overline{Y}\right)d\overline{x} \tag{53}$$

for $\overline{x}_c \equiv x/\delta_c$ and $\overline{\mathcal{L}}_c \equiv \overline{\mathcal{L}}_0\left(\delta_c/\delta_I\right)$. In particular, defining $\delta\omega \equiv \delta\omega_1 + \delta\omega_2$, we have

$$-i\delta\omega_1 \simeq D_m\frac{\Delta'k}{\varepsilon_m^{1/4}}\frac{1}{\mathfrak{I}_0} \sim D_m^{3/4}D_A^{3/4}\frac{k^{3/2}}{L_s^{1/2}}, \tag{54}$$

for $\Delta' \sim k$, and

$$-i\delta\omega_2 \simeq \left|\omega_{*\parallel}^T\right|\frac{\varepsilon_*^{3/4}}{\Delta_{th}^{3/4}}G_0\frac{1}{\mathfrak{I}_0} \sim \left|\omega_{*\parallel}^T\right|\left(\frac{D_m}{D_\perp^e}\right)^{3/4}\left(\frac{D_\parallel^e}{D_A}\right)^{3/4} \tag{55}$$

where

$$\mathfrak{I}_0 \equiv \int d\overline{x}_c\left\{1+\alpha_*^T\overline{x}_c\overline{\mathcal{L}}_c\right\} \equiv \mathfrak{I}_{0R}+i\mathfrak{I}_{0I} \tag{56}$$

and

$$G_0 \equiv \int (1 - \bar{x}\bar{Y}) d\bar{x} . \tag{57}$$

Then we note that most of the contribution to $G_0$ comes from the $\delta_I$ region and, referring to the solution of Eq. (22), we can conclude that this contribution is positive. The same conclusion is valid for $\mathfrak{I}_{0R}$ defined by Eq. (56). In particular, the numerical solution of Eq. (46) leads to find $\mathfrak{I}_{0R} \simeq 2/|\alpha_*^T|^{1/4}$. Likewise, we can verify that $\text{Re}(\delta\omega/\omega_*^{TT}) < 0$. Moreover, if we consider the numerical solution of Eq. (22) we find $G_0 \simeq 2.2$.

Returning to Eq. (20) we note that the term $\bar{x}$ on the r.h.s. can be considered prevalent in the innermost region if, consistently with Eq. (41), $\varepsilon_*^{1/2} > D_m (kL_s)^{1/2} / (\varepsilon_*^{1/4} D_\parallel^e)$, for $\Delta' \sim k$. This implies that $\varepsilon_*^{1/4} (kL_s)^{1/2} < D_\parallel^e / D_A$, a reasonable condition.

## 8. Relevant Comments

We observe that the expression for $\delta\omega$ can be rewritten as

$$-i\delta\omega \simeq D_m \frac{1}{\varepsilon_*^{1/4}} \left[ \left(\frac{k}{L_s}\right)^{1/2} \Delta' + \frac{|\omega_{*\parallel}^T|}{D_A} \frac{1}{\Delta_{th}^{3/4}} G_0 \right] \frac{1}{\mathfrak{I}_0} . \tag{58}$$

Thus the modes, in the considered regime where $\Delta_{th} > \varepsilon_*$, are excited if

$$\left(\frac{k}{L_s}\right)^{1/2} \Delta' > -\frac{|\omega_{*\parallel}^T|}{D_A} \frac{1}{\Delta_{th}^{3/4}} G_0 , \tag{59}$$

that is, even when $\Delta' < 0$ and the basic drift-tearing modes cannot be excited. In fact, we may write a rudimentary extension of Eq. (59) to the case where $\Delta_{th}/\varepsilon_* \to 0$, that is

$$\left(\frac{k}{L_s}\right)^{1/2} \Delta' > \frac{1}{4\Delta_{th}^2 + \varepsilon_*^2} \left[ \varepsilon_*^2 \left(\frac{k}{L_s}\right)^{1/2} \Delta_c' - 4\Delta_{th}^2 \frac{|\omega_{*\parallel}^T|}{D_A \Delta_{th}^{3/4}} G_0 \right], \tag{60}$$

where $\Delta_c'$ is the (considerably high) critical value of $\Delta'$ that can be found for $\Delta_{th} = 0$.

We note that

$$\delta_c \sim \left(\frac{D_m}{D_A}\right)^{1/4} \left(\frac{L_s}{k}\right)^{1/2} \propto \left(\frac{L_s}{B}\right)^{1/2} \left(\frac{n}{kr_p}\right)^{1/4} \frac{1}{T^{1/8}} . \tag{61}$$

Therefore $\delta_c$ decreases quite slowly with temperature and it is clear that the present linearized theory breaks down when the width of the macroscopic magnetic islands that it can produce, as indicated by

$$\delta_{Is} \equiv \left| \frac{\tilde{B}_{x0}}{B'_y k} \right|^{1/2}, \tag{62}$$

becomes the order of $\delta_c$. This corresponds to $\left|\tilde{B}_{x0}\right|/B \sim \varepsilon_*^{1/2}$ that is equivalent $\tilde{\Omega}_{ci}\tilde{\Omega}_{ce} \sim \nu_{ei}\omega_{*\parallel}^T$, where $\tilde{\Omega}_{ci} \equiv e\tilde{B}_{x0}/(m_i c)$ and $\tilde{\Omega}_{ce} \equiv \tilde{\Omega}_{ci} m_i/m_e$. Thus we may conjecture that when this limit is reached the effective value of $\delta_c$ will continue to grow until that of $\delta_I$ is reached and that this will be the island saturation size. It is evident in any case that using the so-called Rutherford model, that is appropriate for the highly resistive tearing mode [2], is not justified for the weakly collisional regimes to which the theory we have presented applies.

**Acknowledgments**

It is a pleasure to thank C. Crabtree and V. Roytershteyn for their collaboration on different parts of the subject treated in this paper. In particular, V. Roytershteyn [12] described first numerically, in his thesis, the transition from the case where $D_\perp^e/D_\parallel^e$ is negligible to the case where it is not. I am grateful to F. Bombarda for her support in bringing this paper to light and to K. Tummel for his numerical solutions.